# DC electricity generation from dynamic polarized water-semiconductor interface

*Yanfei Yan[1,2#], Xu Zhou[3#], Sirui Feng[1,2#], Yanghua Lu[1,2#], Jianhao Qian[4], Panpan Zhang[1], Xutao Yu[1], Yujie Zheng[1], Fengchao Wang[4], Kaihui Liu[3], Shisheng Lin[1,2*]*

1: College of Microelectronics, College of Information Science and Electronic Engineering, Zhejiang University, Hangzhou, 310027, China

2: State Key Laboratory of Modern Optical Instrumentation, Zhejiang University, Hangzhou, 310027, China

3: State Key Lab for Mesoscopic Physics and Frontiers Science Center for Nano-optoelectronics, Collaborative Innovation Center of Quantum Matter, School of Physics, Peking University, Beijing 100871, China

4: CAS Key Laboratory of Mechanical Behavior and Design of Materials Department of Modern Mechanics, CAS Center for Excellence in Nanoscience, University of Science and Technology of China, Hefei, 230027, China

*Corresponding author. Tel: +86-0571-87951555

Email: shishenglin@zju.edu.cn.

**Abstract**

Liquid electricity generator and hydrovoltaic technology have received numerous attentions, which can be divided into horizontal movement generator and vertical movement generator. The horizontal movement generator is limited for powering the integrated and miniaturized energy chip as the current output direction is depending on the moving direction of the water droplet, which means a sustainable and continuous direct current output can be hardly achieved because of the film of limited length. On the other hand, the existing vertical movement generators include triboelectricity or humidity gradient-based liquid electricity generator, where the liquid or water resource must be sustainably supplied to ensure continuous current output. Herein, we have designed an integratable vertical generator by sandwiching water droplets with semiconductor and metal, such as graphene or aluminum. This generator, named as polarized liquid molecular generator (PLMG), directly converts the lateral kinetic energy of water droplet into vertical direct-electricity with an output voltage of up to ~1.0 V from the dynamic water-semiconductor interface. The fundamental discovery of PLMG is related to the non-symmetric structure of liquid molecules, such as water and alcohols, which can be polarized under the guidance of built-in field caused by the Fermi level difference between metal and semiconductor, while the symmetric liquid molecules cannot produce any electricity on the opposite. Integratable PLMG with a large output power of ~90 nW and voltage of ~2.7 V has been demonstrated, meanwhile its small internal resistance of ~250 kΩ takes a huge advantage in resistance matching with the impedance of electron components. The PLMG shows potential application value in the Internet of Things after proper miniaturization and integration.

**Main**

Water covers the ~71% surface of the earth and there are many ways of producing electricity from water[1-6], where the most popular and commercialized way is driving the electromagnetic generator through the movement of water[7, 8] since the Faraday effect was revealed in the middle of 19th century[9]. In the past decade, generating electricity from the moving water in the absence of a magnetic field has received numerous attentions as a result of the requirement of portable energy sources for highly developed human society, especially the Internet of Things (IoTs). There are two lines of utilizing the mechanical energy of water: one is the vertical movement of water, such as triboelectricity[10-12], contact electrification[13,14] and moisture gradient[15,16] induced electricity, where the water droplet needs to sustainably hit on the surface of the device or a humidity gradient needs to be reestablished. And the other is named horizontal movement generator[17-21] where the water droplet moves or water is evaporated along the surface of graphene or functional group decorated carbon film. For the vertical movement type, the recollection process of water adds complexity to the design of the generator with required and sustainable voltage or current. For the lateral movement type, the output electricity direction is dependent on the moving direction of the water droplet, which means a sustainable electricity output is hardly achieved as the length of the film is limited.

Learning from the information society, every chip is composed of numerous transistors, an integratable liquid electricity generator is indispensable for the sustainable advancement of energy density, which requires a liquid generator sustainably worked under a circular space without the evaporation of liquid or water droplet.[22-27] Only based on this kind of generator, the miniaturization of the water generator for constructing a water generator chip different from the electromagnetic generator can be promised. For the direct current output, we must design a generator whose electricity output direction is not changed as the moving direction of water droplet changes, which means the traditional lateral water generator cannot be utilized for liquid generator energy chip. For the sustainable electricity output, we must design a closed space without evaporating the water as it takes energy to recover the water resource or humidity gradient. Herein, we designed a water droplet sandwiched dynamic polarized water generator where the water droplet is confined between metal (such as graphene and aluminum) and semiconductor substrates. The key to the electricity output is the polarization of water and other nonsymmetric molecular on the semiconductor substrate as a result of Fermi level difference between metal and semiconductor. The direction of electricity generation is fixed while the moving direction of liquid changes, that is, a direct current with fixed direction can be generated when we move the water droplet between the metal and semiconductor at will. As one of the typical PLMG, the dynamic graphene/water/n-silicon and aluminum (Al)/water/n-silicon generator with a small moving water droplet output a large direct-voltage of ~0.28 and ~1.0 V, respectively. In addition, other features including the small internal resistance of ~250 kΩ, the large output current of ~2.2 μA and power of ~90 nW, declare its improvement on performance in the field of the self-powered generators. The performance of the generator is further improved to a voltage of ~2.7 V by connecting 3 generators in-series, which is already comparable with a commercial dry battery. The demonstrated generator with novel mechanism has the advantages of realizing repeatable, low cost and integratable PLMG based energy chip matching impedance with electron components. In addition, the durability and continuity of the dynamic process of the heterostructure generator also broaden its application in the field of bioenergy harvesting and flexible devices.

The setup of the device structure and measurement process is shown in Figure 1a. The PLMG consists of a polished n-typed silicon substrate (72 mm × 42 mm) with Ti/Au metal films electrode deposited on back, a monolayer graphene film attached on polyethylene terephthalate (PET) plate and a moving pure water droplet (deionized water with ion concentration less than $10^{-18}$ mol/L) in the gap (depth ≈ 1 mm, length ≈ 72 mm and width ≈ 8 mm) between them (Figure 1a). The short-circuit current and open-circuit voltage are collected separately. Meanwhile, a high-speed camera recorded the motion of water droplet, from which one can extract the droplet speed. From top to bottom of the plate, each movement of the droplet produces one current and voltage peak at a time. The dependence of voltage/current and speed is shown in Figure 1b and Figure 1c, respectively. Notably, whatever direction of the moving droplet is, a positive current and voltage from graphene to silicon are collected, which illustrates that the designed PLMG is based on a direct-electricity principle. Such a feature of direct-electricity output is of great significance for developing liquid generator integrated energy chip for powering distributed sensors. In the water-based PLMG, direct electricity is generated from a dynamic polarized water molecular-semiconductor interface. As the first principle simulation shows, the water molecular placed horizontally (Figure 1d) between graphene and silicon relax to a stable state after a period of time (Figure 1e), and the water molecule rotate with oxygen atom end and hydrogen atom end of the water molecule pointing to the graphene and silicon, respectively. Figure 1f shows the simulated and calculated distribution of charge density by Bader Charge Analysis method[28-31], where negatively-charged oxygen atoms induce positively-charged holes ($-1.5 \times 10^{-4}$ e·Å$^{-3}$ charge density) at the interface water/graphene meanwhile positively charged hydrogen atoms induce negatively-charged electrons ($1.5 \times 10^{-4}$ e·Å$^{-3}$ charge density) at the interface water/silicon. In this way, the water molecules are polarized by the Fermi level difference ($\Delta E_F$) of two plates and induce the charges accumulation in two separate plates. A complete electricity generation process in water-based PLMG is illustrated in Figure 3g. When the water droplet is not contacting the graphene and n-silicon, the polarization direction of water molecules is randomly aligned. When the water droplet contacts with graphene and n-type silicon, the water molecules are polarized and aligned. As the water droplet moves, the water molecular attracted holes leave the graphene layer and recombine with the electrons escaping from the n-type silicon substrate. It is noteworthy that the water droplet can conduct electricity, which may be a result of contact-induced carrier injection as revealed in the contact electrification process at the liquid-solid interface[6, 7]. According to the first principle calculations, the Fermi level difference between graphene or metal and semiconductor is crucial for developing the aligned and polarized water molecules.

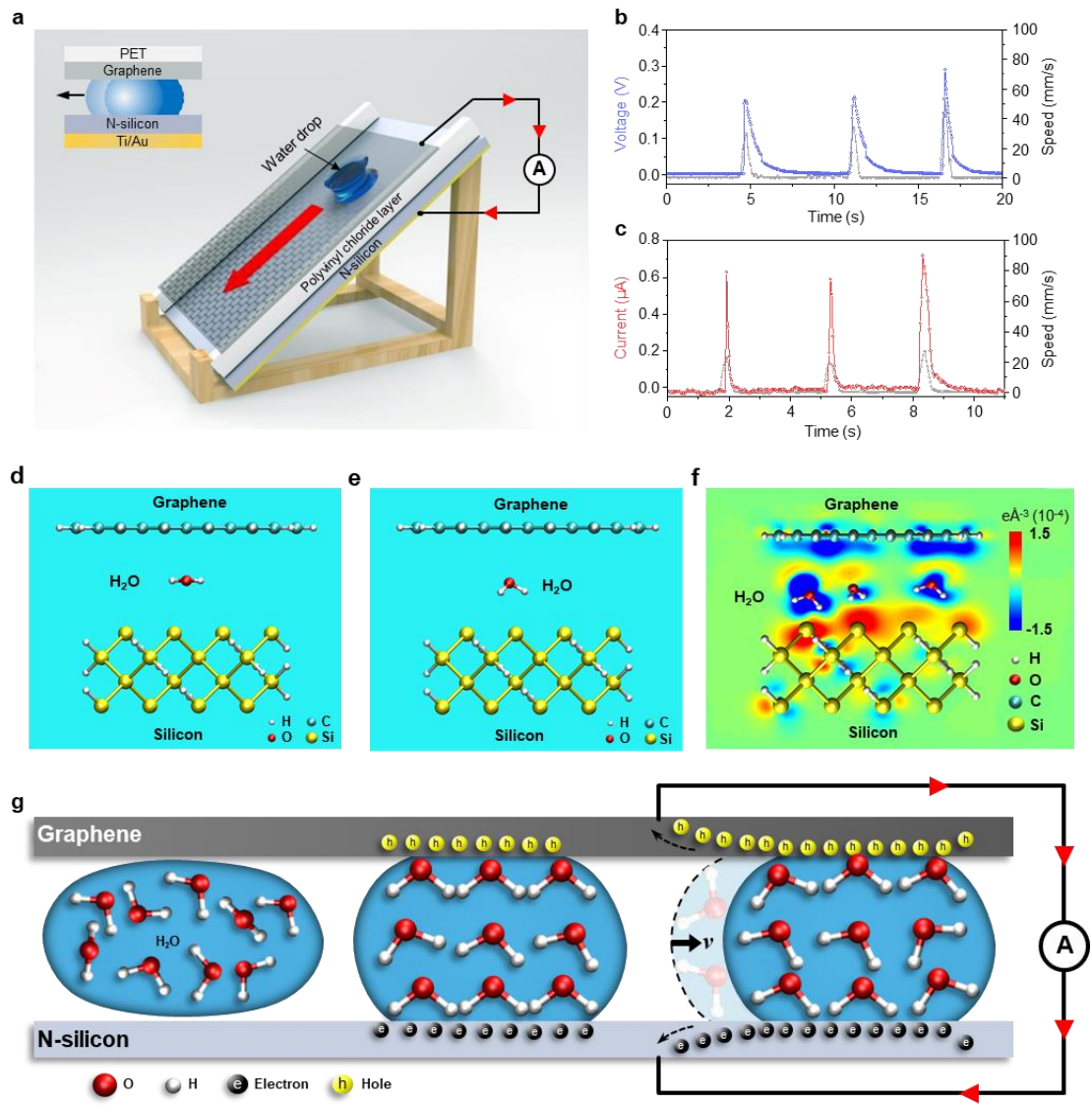

**Figure 1. The schematic of dynamic graphene/water/semiconductor direct-electricity generator. a)** the dynamic generator consists of a polished n-typed silicon substrate (72 mm × 42 mm) with Ti/Au metal films deposited on back, a monolayer graphene film attached on thin polyethylene terephthalate (PET) plate, two polyvinyl chloride strips as an insulator layer between them and a moving pure water droplet in the gap (depth ≈ 1 mm, length ≈ 70 mm and width ≈ 8 mm). A current/voltage meter connects the electrodes of graphene and n-silicon with two leads. The inset figure is the side view. **b), c),** the voltage

**generation peaks (b, blue line), current generation peaks (c, red line) and speed peaks (yellow line) dependent on the time. Three peaks represent three measurements where water droplet moving from top to bottom producing one peak. d), e), a density functional theory simulation was carried in which the water molecules placed horizontally (d) between graphene and silicon plates slowly rotate until the hydrogen atom end toward (e) silicon after the energy relaxation process. White: H; Cyan: C; Red: O; yellow: Si. f, the calculated electron density difference distribution of graphene/water/ silicon generator. The electron depletion and electron accumulation represent in blue and red, respectively. Silicon layers obtain 0.15 electrons when hydrogen atoms point toward silicon. g, The schematic illustration of the electricity dynamic generation process in generator. Water molecules are chaotically arranged before droplet contacts the graphene and n-silicon plates, as illustrated in left droplet. Molecules are polarized when droplet contacts two plates, as shown in middle droplet. The right droplet shows that water droplet moves to right direction with speed of v, and electrons and holes escape from the bondage of polarization to flow into external circuit.**

Experimentally, as shown in Figure 2a and 2b, the open-circuit voltage ($V_{oc}$) and short-circuit current ($I_{sc}$) of the generator both indicate a positive nonlinear correlation relationship with the droplet speed of below 15 mm/s. With the droplet speeds up, more charges are generated from the water/semiconductor interface, leading to the enhancement of the current and voltage. When the droplet speed continues to increase beyond 15 mm/s with the droplet volume of 30 μL, the $V_{oc}$ and $I_{sc}$ both reach the saturation of 0.28 V and 0.80 μA, respectively. Such saturation behavior of $V_{oc}$ or $I_{sc}$ originates from and is also limited by the Fermi level difference ($\Delta E_F$) between two materials in this generator, which is the driving force to induce the polarization of water molecules and the accumulation of electrons or holes in plates. From the perspective of the molecular polarization process, the $V_{oc}$ and $I_{sc}$ reaching saturation are resulted from the degree of water "depolarization" reaches saturation when droplet moves at high speed. The experiment data of $V_{oc}$ and $I_{sc}$ dependence on the droplet speed $v$ can be fitted well as an exponential function with voltage/current saturation of 0.28 V/0.80 μA, $V_{oc} = -0.25 \times e^{-0.15v} + 0.28$ and $I_{sc} = -0.69 \times e^{-0.16v} + 0.80$ (0.28 and 0.80 is the saturation voltage and displacement current, respectively), which is fully consistent with the mathematical prediction of equation (1). When the droplet moves beyond the speed threshold of ~40 mm/s, the $V_{oc}$ saturation keeps constant of around 0.28 ± 0.03 V with different droplet volumes, which is limited by the $\Delta E_F$ according to the electrodynamic principle (Figure 2c). However, the measured current $I_{sc}$ nonlinearly increases with bigger droplet volume and finally reaches a saturation value of ~2.2 μA with volume beyond 100 μL (Figure 2d). This trend of growing current $I_{sc}$ is actually due to the decreasing of the internal resistance of graphene/water/silicon generator. The total internal resistance $R'_{internal} = V_{oc}/I_{sc}$ (Supplementary Figure 1) calculated from the curves where $V_{oc}$ (Figure 2c) and $I_{sc}$ (Figure 2d) saturate with the increase of droplet volume is well coincident with the experimental curve of internal resistance, accounting for dependence of voltage/current on varied resistance and droplet volume. The internal resistance can consume electricity and reduce the output power of the generator, so it is important to match the resistance of the electronic device to the generator. The dependence of load voltage and current on load resistance is shown in Figure 2e. With the load resistance increases, the load voltage increases and reaches a maximum of 0.28 V while the load current with a maximum of 0.80 μA decreases to nearly zero, on the condition of droplet speed of ~40 mm/s and droplet volume of ~30 μL. Accordingly, the corresponding power output of the generator as a function of electrical load resistance was calculated in Figure 2f, where the load resistance of ~250 kΩ corresponds to the power maximum of ~90 nW indicates the equivalent internal resistance of this dynamic generator. Such low internal resistance in an order of magnitude of kilohm is well matching with the impedance of those semiconductor-based information electronic devices, which can save much energy loss and output maximum power and improve the efficient utilization of the mechanical energy of water droplets. Furthermore, this dynamic generator with high output power can directly convert the kinetic energy of water droplet to direct-electricity with the high conversion efficiency of 6.1‰ (the detail calculation is proposed in method part).

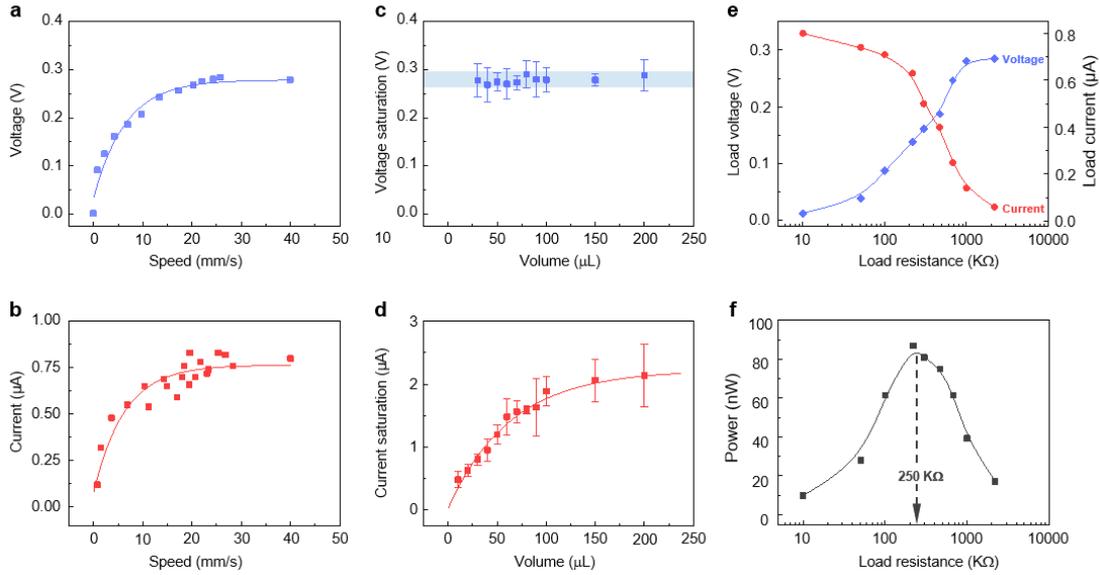

**Figure 2. The performance characterization of the dynamic graphene/water/silicon generator. a) and b), the speed dependence of open-circuit voltage $V_{oc}$ (a) and short-circuit current $I_{sc}$ (b) of the dynamic graphene/water/silicon generator with the water volume of 30 μL, respectively. c), d), the relationship of the water volume and the open-circuit voltage saturation (c) and short-circuit current saturation (d) of this generator at large droplet speed of ~40 mm/s, respectively. The colour strip in blue shows the average voltage saturation with error of 0.28 ± 0.03 V. e, the load voltage (blue) and load current (red) output as a function of electrical load resistance. f), the dependence of output power of the generator on the electrical load resistance. Correspond to the output power maximum, generator internal resistance of ~250 kΩ is indicated by the dotted arrow. All the spots in figures are experimental data and solid curves are the fitting lines.**

Besides the direct current output, the voltage output can be accurately designed through managing the $\Delta E_F$ between two plates which encapsulate the water droplets. Keeping the n-silicon with a work function of 4.34 eV unchanged, the conducting materials with different work functions[32-34] include Au (5.10 eV), graphene (4.60 eV), Cu (4.48 eV), Ag (4.26 eV), Pb (4.25 eV), and Al (4.28 eV) have been used to generate electricity. The bandgap alignment between those conducting materials has been shown in Figure 3a. As shown in Figure 3b, the maximum output voltage of those PLMG generators at droplet speed of 40 mm/s and volume of 30 μL are 0.30/0.28/0.11/-0.08/-0.12/-0.99 V, respectively, while the $\Delta E_F$ between these metals and silicon is about 0.76/0.26/0.14/-0.08/-0.09/-0.06 eV. By comparing the voltage value and the $\Delta E_F$ between these metals and n-silicon, we can find that the direction of voltage depends on the sign of $\Delta E_F$, where the positive current is defined as the positive charge flowing from the metal to the silicon in the external circuit. And the voltage output is generally proportional to the absolute value of $|\Delta E_F|$, but the existence of aluminum oxide film on the surface of aluminum causes a reduction in the work function and consequently an increase of voltage output. Furthermore, with the advantage of low cost and easy-availability, such a generator with an aluminum plate is perspective on massive production. The above experiments not only verify the decisive influence of the magnitude and direction of the Fermi level on the size and direction of power generation but also prove the feasibility of PLMG for designed voltage output.

To strength the conclusion of PLMG is truly related with the polarization of water and other nonelectrolyte liquids with asymmetric atomic structure, we replaced the water with a non-polar solution such as carbon tetrachloride ($CCl_4$), normal hexane ($C_6H_{14}$), and other polar solutions ethanol ($C_2H_5OH$) and methanol ($CH_3OH$). When 30 μL liquid slides between graphene and n-silicon at a speed of 40 mm/s, a polar solution such as $C_2H_5OH$ and $CH_3OH$ can generate direct current with voltages of 0.65 V and 0.60 V, respectively. However, those non-polar molecules ($CCl_4$ and $C_6H_{14}$) generate no voltage (Figure 3c), indicating that the polarization process is essential in power generation. This phenomenon strongly suggests that only polar nonelectrolyte molecules rather than non-polar electrolyte molecules can generate electricity from the liquid/semiconductor interface, of which water is the most suitable choice due to its non-pollution and abundant storage. Besides, by adding sodium chloride (NaCl) to the water with the increasing ions concentration from 0, 0.1, 0.5 to 1.0 mol/L, the voltage output decreases from 0.28, 0.13, 0.10 to 0.05 V (Figure 3d). On one hand, the existed $Cl^-$ anions and $Na^+$ anions are more likely to be adsorbed on the water/silicon interface compared with water/graphene interface, while $Cl^-$ is more preferentially adsorbed at the water/silicon interface than $Na^+$ cations, which induces opposite charges in silicon (or graphene) reducing contribution of the water to charge accumulation (Supplementary Figure 2). On the other hand, the addition of ions interrupts and breaks the polarization of some water molecules. Taking chloride ions as an example, in water molecules, negatively charged chloride ions attract positively charged hydrogen atoms. This phenomenon changes the polarization of water molecules and prevents the polarization of water molecules from being reduced, which further reduces voltage/current output (Supplementary Figure 3).

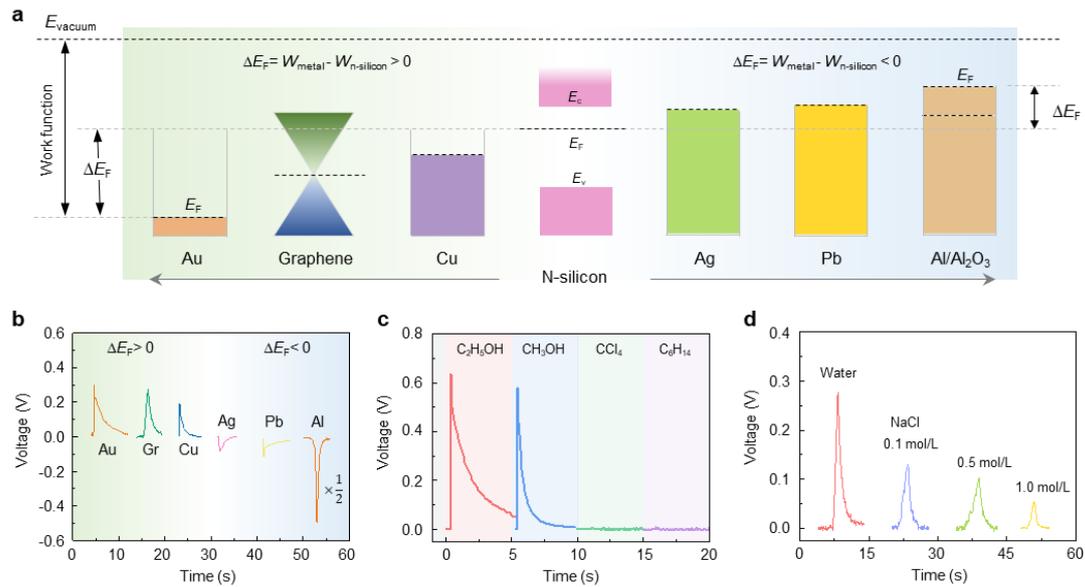

**Figure 3. The physical mechanism based on the Fermi level difference and molecules polarization. a)** The schematic diagram of simplified electronic band alignment of different metals and n-silicon. The long grey dash line and short black dash lines present the vacuum energy level $E_{vacuum}$ and Fermi level $E_F$ of different materials, respectively. $\Delta E_F$ shows the fermi level difference between n-silicon and different metals. **b)** The output voltage curves of metal/water/silicon generator with different metal plates. The green and blue sections of background in (a) and (b) represent the $\Delta E_F > 0$ and $\Delta E_F < 0$, respectively. **c)** The output voltage of dynamic graphene/liquid/silicon generator with different liquids, including $C_2H_5OH$, $CH_3OH$, $CCl_4$ and $C_6H_{14}$. **d)** The output voltage curves of dynamic graphene/NaCl solution/silicon generator with different NaCl concentration from 0-1.0 mol/L. All the experiment carried out at droplet speed of 40 mm/s and droplet volume of 30 μL.

Although a direct-voltage of up to ~1.0 V can be generated by moving a small drop of water in the PLMG, the ability to connect these PLMGs is crucial to sustainably support the electrical energy of distributed sensors in the age of IoTs. As illustrated in Figure 4a and 4b, a PLMG consists of a silicon wafer (~5 cm in diameter) glued to the bottom of a plastic container and a small copper (~4 cm in diameter) stuck on the center of the plastic cap covering the container. Deionized water was injected into the container sealed with glue, forming a vertical movement generator. By continuously shaking the generator and sliding water droplets between aluminum and n-silicon, this PLMG can continuously output an average of ~0.90 V of $V_{oc}$ and ~0.65 μA of $I_{sc}$, respectively (Figure 4c). Importantly, the output voltage can be easily boosted by simply connecting such generators in series like "building block", as shown in Figure 4d, where 3 generators in series can generate sustained direct-voltage of ~2.7 V, which is already comparable to a commercial dry battery voltage of 1.5 V. These demos of the dynamic vertical movement generator show compelling evidence of the excellent voltage output performance and potential in fabricating the liquid generator energy chip after proper miniaturization and integration.

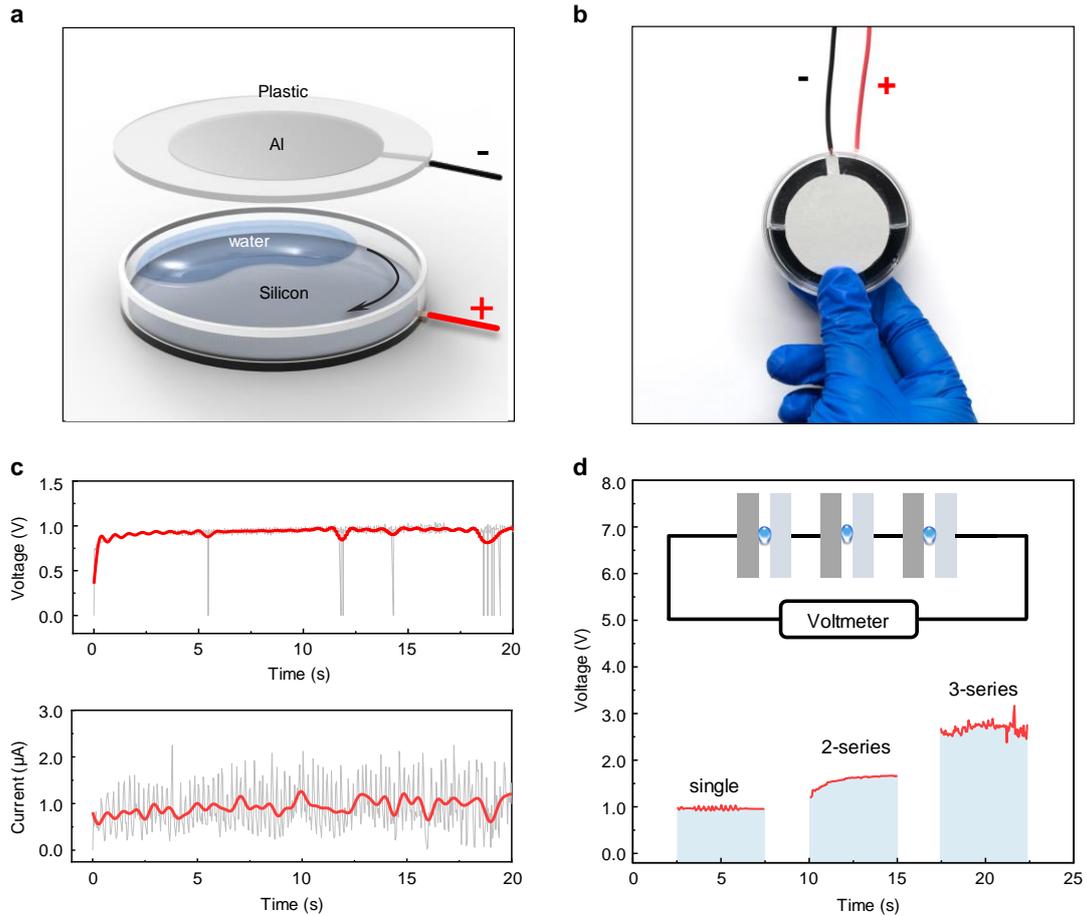

**Figure 4. An application demo of the dynamic metal/water/semiconductor generator.** The illustration (a) and the real device picture (b) of a disc-shaped generator consists of a n-silicon wafer glued on the bottom of a plastic boat (~5 cm in diameter) and a small aluminium (~4 cm in diameter) stuck on the center of the plastic cap covering the boat with deionized water in the boat sealed by glue. By continuously shaking generator and sliding water droplet between copper and silicon, this simple generator can continuously output electricity through the two leads connecting electrodes respectively of the metal and silicon with plus sign and minus sign indicating current output direction. c) The **continuous output voltage (upper panel) and current (bottom panel) curves dependent on the time when one keeps shaking the generator above. The red curve represents the average value of experimental data curve in gray by smoothing. d) The voltage output curves of single, 2 and 3 generators in series. The insert is the schematic diagram of three dynamic generators in-series.**

In summary, we have demonstrated a metal/liquid/semiconductor direct-electricity generator that can generate direct current from a dynamic liquid-semiconductor interface by moving a small drop of liquid which resulting from the "appear-disappear" of polarized liquid molecules. . An excellent direct-electricity output of voltage/current up to ~1.0 V/ ~2.2 μA has been achieved, in which output voltage can be designed through changing the $\Delta E_F$ between different metal and semiconductor plates. This PLMG with liquid sealed in its vertical structure can continuously generate electricity for powering various distributed portable devices, which can be finely integrated for an improved voltage output up to ~ 2.7 V. All those features together with its low internal resistance make it a strong candidate for integrated energy chips. The PLMG offers a competitive and feasible solution for the fast developed IoTs, where numerous widely distributed needs to be powered in-situ in a dark environment with available mechanical energy.

## METHODS

**Devices fabrication and experimental data acquisition.** The double side polished n-type silicon wafer was removed the oxide layer by dipping it in 10 wt% HCl for 10 minutes. Then Ti/Au (20 nm/50 nm) electrode was deposited with magnetron sputtering on back side of silicon wafer. The graphene film by chemical vapor deposition growth was transferred on a polyethylene terephthalate substrate. The two plates were fixed together and separated by an electrical insulator called polyvinyl chloride with a gap (length $\times$ width $\times$ depth $\approx$ 72 mm $\times$8 mm $\times$1 mm). The speed of water droplets is extracted by an image processing analysis software Image-Pro Plus (American MEDIA CYBERNETICS), where moving water drop was recorded by a high-speed camera (US Phantom, VEO 710). The voltage and current data were collected by a voltmeter with sampling rate of 25 s$^{-1}$ (Keithley 2010) and a Pico ammeter with sampling rate of 100 s$^{-1}$ (Keithley 6485), respectively. The power conversion efficiency of the dynamic conductor/water/semiconductor generator is calculated with the power output and potential energy transformation. The water droplet with the volume of 30 μL was free fall in 1.07 seconds accelerating from a slope with 60 degree with height of 52 mm. And the droplet speed is 43 mm/s extracted by software. The droplet kinetic energy of 2.34$\times$10$^{-8}$ J is partially converted from the droplet potential energy (i.e. the energy loss) of 1.53$\times$10$^{-5}$ J, while the output electric energy is 86.80 nW, corresponding to the generating energy of 9.29$\times$10$^{-8}$ J. Therefore, the energy conversion efficiency of the generator is about 6.1‰, which is calculated with the ratio of generating energy and the energy loss.

**The work function of n-silicon.** The Fermi level of the semiconductor can be calculated by the formula below as below:

$$W_{n-silion} \approx E_i + k_B T \ln\frac{N_D - N_A}{n_i}$$

where $E_i$ is the middle value of the band gap, $k_B$ is the Boltzmann constant, $T$ is the temperature, $n$ is the electron concentration and $p$ is the hole concentration. The $n_i$ is the intrinsic carrier concentration of the semiconductor. The electron concentration and intrinsic carrier concentration of the n-type Si substrate used here is $4.63 \times 10^{14}$ cm$^{-3}$ and $1.50 \times 10^{10}$ cm$^{-3}$, respectively. Therefore, the work function of the n-type Si is calculated as 4.34 eV based on the above equation.

**The electron density difference.** The simulation system consists of a monolayer graphene, four layers of silicon atoms, and water molecules. Dangling bonds at the edges of graphene and silicon were passivated by hydrogen atoms. The electron density was calculated using the program Gaussian with B3LYP functional. The def2SVPP and def2SVP basis set were used for substrates (graphene and silicon) and water, respectively. Meanwhile, dispersion correction DFT-D3 (BJ) was added. To obtain the electron density difference, the electron density contours for substrate (silicon and graphene), water, and the whole system were calculated individually. Then the electron density difference was generated by subtracting substrate's and water's electron density from the whole system. Multiwfn program was used to process the grid data to the electron density.

## Data availability

The data that support the findings of this study are available from the corresponding author upon reasonable request.

**Acknowledgement**

S. S. Lin thanks the support from the National Natural Science Foundation of China (No. 51202216, 51502264, 61774135 and 51991342) and Special Foundation of Young Professor of Zhejiang University (Grant No. 2013QNA5007), Beijing Natural Science Foundation (JQ19004), Beijing Excellent Talents Training Support (2017000026833ZK11), Bureau of Industry and Information Technology of Shenzhen (No. 201901161512), Key-Area Research and Development Program of Guangdong Province (Grant No. 2019B010931001, 2020B010189001). The numerical calculations have been performed on the supercomputing system in the Supercomputing Center of University of Science and Technology of China. Project funded by China Postdoctoral Science Foundation (2019M660001) and Postdoctoral Innovative Personnel Support Program (BX20180013).

**Author Contributions:** S. S. Lin designed the experiments, analysed the data and conceived all the works. Y. F. Yan, X. Zhou, S. R. Feng and Y. H. Lu carried out the experiments, discussed the results and wrote the paper. J. H. Qian, P. P. Zhang, X. T. Yu, Y. J. Zheng discussed the results and assisted with experiments. F. C. Wang and K. H. Liu discussed the results, analysed the data and wrote the paper. All authors contributed to the scientific discussion and the writing of the paper.

**Competing Interests:** The authors declare no competing financial interest. Readers are welcome to comment on the online version of the paper.

**Supplementary Information** is Available in the online website.